\documentclass[aps,prd,twoside,twocolumn,nofootinbib,10pt,showpacs,floatfix,longbibliography]{revtex4-1}
\usepackage{amsmath,amssymb}
\usepackage{graphicx,bm}
\usepackage{graphicx}
\usepackage{slashed}
\usepackage{epstopdf}
\usepackage[usenames]{color}
\usepackage{float}
\usepackage{hyperref}
\usepackage{subcaption}
\usepackage{rotating}
\usepackage{color,xcolor}
\usepackage{multirow}
\usepackage{dcolumn}
\usepackage{overpic}
\usepackage{lipsum}
\usepackage{booktabs}
\usepackage{makecell}
\usepackage{todonotes}
\usepackage{diagbox}
\usepackage{array}
\newcolumntype{|}{!{\vline}}

\newsavebox{\tablebox}
 
\begin{document}
\title{Scattering of $\Lambda_{c}\Lambda_{c}$ and $\Lambda_{c}\bar{\Lambda}_{c}$ in chiral effective field theory}

\author{Zhe Liu$^{1,2,5}$}\email{zhliu20@lzu.edu.cn}
\author{Hao Xu$^{4}$}\email{xuh2020@nwnu.edu.cn}
\author{Zhan-Wei Liu$^{1,2,3,5}$}
\author{Xiang Liu$^{1,2,3,5}$}\email{xiangliu@lzu.edu.cn}
\affiliation{
$^1$School of Physical Science and Technology, Lanzhou University, Lanzhou 730000, China\\
$^2$Lanzhou Center for Theoretical Physics,
Key Laboratory of Theoretical Physics of Gansu Province,
Key Laboratory of Quantum Theory and Applications of MoE,
Gansu Provincial Research Center for Basic Disciplines of Quantum Physics, Lanzhou University, Lanzhou 730000, China\\
$^3$MoE Frontiers Science Center for Rare Isotopes, Lanzhou University, Lanzhou 730000, China\\
$^4$Institute of Theoretical Physics, College of Physics and Electronic Engineering,
Northwest Normal University, Lanzhou 730070, China\\
$^5$Research Center for Hadron and CSR Physics, Lanzhou University and Institute of Modern Physics of CAS, Lanzhou 730000, China}

\date{\today}
\begin{abstract}

We investigate the $S$-wave scatterings of $\Lambda_c\Lambda_c$ and $\Lambda_{c}\bar \Lambda_{c}$ systems within a unified chiral effective field theory framework up to next-to-leading order. The contact low-energy coupling constants are determined by fitting to the lattice QCD results for the $\Lambda_c\Lambda_c$ scattering phase shift at an unphysical pion mass. After extrapolating to the physical pion mass, we find a repulsive interaction in the $I(J^P)=0(0^{+})$ $\Lambda_c\Lambda_c$ channel, consistent with the lattice QCD simulation. On the $\Lambda_{c}\bar \Lambda_{c}$ side, using the fitted contact low-energy constants, we predict the phase shifts and potentials for $\Lambda_c \bar\Lambda_c$ scattering in the $I(J^{PC})=0(0^{-+})$ and $0(1^{--})$ channels. Attractive interactions are found in both channels, each allowing for the formation of bound states. In particular, the attraction in the $0(1^{--})$ $\Lambda_c \bar\Lambda_c$ channel is stronger. In addition, our analysis reveals that the spin-spin term caused by the two-pion exchange contributes significantly to the interactions, leading to a distinct mass splitting between the $0(0^{-+})$ and $0(1^{--})$ $\Lambda_c \bar\Lambda_c$ channels. 

\end{abstract}

\maketitle
\thispagestyle{empty} %
\section{Introduction}

In hadron physics, the study of dibaryon (or hexaquark) states has a long and intricate history \cite{Clement:2016vnl}.
Inspired by studies of the nuclear force, the interactions of dibaryon systems have been extensively analyzed using various theoretical methods \cite{Haidenbauer:2013oca, Maessen:1989sx, Reuber:1993ip, Rijken:1998yy, Korpa:2001au, Haidenbauer:2015zqb, Li:2016mln, Huang:2019esu, Song:2020isu, Song:2021war, Liu:2021pdu, Alcaraz-Pelegrina:2022fsi, Liu:2022nec, Sekihara:2023ihc}.
Meanwhile, in recent years, the HAL QCD method provides an approach for deriving baryon-baryon interaction potentials from first principles \cite{HALQCD:2014okw, Nemura:2014kea, Sasaki:2015ifa, Miyamoto:2017ynx, HALQCD:2018qyu, HALQCD:2019wsz, Inoue:2021tdt}.

Motivated by the aforementioned studies of light dibaryons, it is natural to extend the investigation to dibaryons containing heavy quarks, particularly two-heavy-baryon systems. The combination of chiral symmetry and heavy quark symmetry makes heavy hadron systems ideal platforms for studying hadron interactions. Among these, the interactions of the $\Lambda_{c}\Lambda_{c}$ and $\Lambda_{c}\bar{\Lambda}_{c}$ systems have been investigated using various theoretical frameworks. Lee \textit{et al.} systematically studied heavy dibaryon and heavy baryonium systems within the one-boson-exchange model \cite{Lee:2011rka}. They concluded that there are no $H$-dibaryon-like states in the $\Lambda_{c}{\Lambda}_{c}$ sector but that $\Lambda_{c}\bar{\Lambda}_{c}$ bound states can exist. In contrast, Refs.~\cite{Meguro:2011nr, Liu:2012zzo} suggested the possible existence of a double-charm $\Lambda_{c}{\Lambda}_{c}$ bound state. Chen \textit{et al.} investigated the $\Lambda_{c}{\Lambda}_{c}$ interaction in the framework of chiral effective theory (ChEFT) \cite{Chen:2022iil}.
In addition, various other methods have been applied to study double-charm dibaryon systems \cite{Leandri:1997ge, Julia-Diaz:2004ict, Froemel:2004ea, Li:2012bt, Xia:2021tof, Carames:2015sya, Vijande:2016nzk, Lu:2017dvm, Chen:2017vai, Garcilazo:2020acl, Ling:2021asz, Junnarkar:2022yak, Liu:2022rzu, Cao:2024hnn}. For instance, in Ref.~\cite{Huang:2013rla}, Huang, Ping, and Wang studied $H$-like dibaryon systems with heavy quarks within the quark delocalization color screening model, and their results indicated that the $\Lambda_{c}{\Lambda}_{c}$ interaction is repulsive.
On the other hand, several theoretical studies have focused on the $\Lambda_{c}\bar{\Lambda}_{c}$ interaction or hidden-charm hexaquark states \cite{Gerasyuta:2013esc, Chen:2016ymy, Wan:2019ake, Wang:2021qmn}. In Ref.~\cite{Song:2022yfr}, Song and collaborators studied possible molecular states arising from the $\Lambda_{c}\bar{\Lambda}_{c}$ interaction mediated by $\omega$ and $\sigma$ exchanges. They found two bound states with quantum numbers $J^{PC}=0^{-+}$ and $1^{--}$ and discussed their production in nucleon-antinucleon collisions.

On the experimental side, the $\Lambda_{c} \bar{\Lambda}_{c}$ system has attracted considerable interest. The Belle Collaboration first reported the $Y(4630)$ resonance, with mass $M=4634^{+8}_{-7}(\text{stat})^{+5}_{-8}(\text{sys})$ MeV and total width $\Gamma_{\text{tot}}=92^{+40}_{-24}(\text{stat})^{+10}_{-21}(\text{sys})$ MeV, in the cross section of $e^+e^- \rightarrow \Lambda_{c} \bar{\Lambda}_{c}$ \cite{Belle:2008xmh}.
More recently, the BESIII Collaboration measured the cross sections for this process \cite{BESIII:2017kqg, BESIII:2023rwv}. They found no indication of the $Y(4630)$ resonant structure reported by Belle. Additionally, they searched for $H_{c}^{\pm}$, a possible $\Lambda_{c} \bar{\Sigma}_{c}$ bound state, via the $e^+e^- \rightarrow \pi^{+}\pi^{-}\Lambda^{+}_{c} \bar{\Lambda}^{-}_{c}$ process, but observed no statistically significant signal \cite{BESIII:2025zgc}.

This leads to a natural question: What is the relation between the $\Lambda_{c} \bar{\Lambda}_{c}$ interaction and the observed charmoniumlike states?
In Refs.~\cite{Cao:2019wwt, Guo:2024pti}, final state interactions were introduced to explain the pronounced enhancements around the $\Lambda_{c}\bar{\Lambda}_{c}$ threshold in the BESIII data~\cite{BESIII:2017kqg, BESIII:2023rwv}. In Ref.~\cite{Milstein:2022bfg}, the mixing of $S$-wave and $D$-wave components of the $\Lambda_{c}\bar{\Lambda}_{c}$ wave function due to the tensor part of the final state interactions was considered to explain the near-threshold energy dependence of the $e^+e^- \rightarrow \Lambda_{c} \bar{\Lambda}_{c}$ cross section observed by the Belle and BESIII Collaborations~\cite{Belle:2008xmh, BESIII:2017kqg}.
Furthermore, the $\Lambda_{c}\bar{\Lambda}_{c}$ structure has also been considered in studies of the production and decay of the $Y(4260)$ \cite{Qiao:2005av, Qiao:2007ce, Chen:2011cta}.
A systematic investigation of the $\Lambda_{c}\bar{\Lambda}_{c}$, $\Sigma_{c}\bar{\Sigma}_{c}$, and $\Lambda_{b}\bar{\Lambda}_{b}$ interactions was carried out in Ref.~\cite{Chen:2013sba}, where the $Y(4260)$ and $Y(4360)$ were interpreted as possible $\Lambda_{c}\bar{\Lambda}_{c}$ baryonia. Several studies based on direct analyses of experimental data have investigated the $\Lambda_c \bar{\Lambda}_c$ molecular state~\cite{Dong:2021juy, Salnikov:2023qnn, Nakamura:2023obk}.

In summary, it is evident that studies of exotic and charmoniumlike states can benefit significantly from investigations of the $\Lambda_{c}\Lambda_{c}$ and $\Lambda_{c}\bar{\Lambda}_{c}$ interactions. Moreover, given the intrinsic relation between the $\Lambda_{c}\Lambda_{c}$ and $\Lambda_{c}\bar{\Lambda}_{c}$ systems, it is desirable to describe both within a unified framework, ideally with a common set of parameters.

In this work, we therefore present a simultaneous study of the $S$-wave $\Lambda_{c}\Lambda_{c}$ and $\Lambda_c \bar\Lambda_c$ interactions using ChEFT in the heavy hadron formalism. Studies of low-energy baryon-baryon interactions have achieved considerable success within the ChEFT framework, e.g., for nucleon-nucleon systems \cite{Epelbaum:2008ga, Machleidt:2011zz, Holt:2014hma, Machleidt:2016vlh, Epelbaum:2019kcf, Hammer:2019poc} and hyperon-nucleon systems \cite{Haidenbauer:2011za, Haidenbauer:2019boi, Petschauer:2020urh, Gerstung:2020fzk}. By incorporating heavy quark symmetry, ChEFT has also demonstrated its effectiveness in describing the low-energy dynamics of two heavy hadrons \cite{Chen:2022iil, Wang:2018atz, Meng:2019ilv, Wang:2019ato, Xu:2021vsi, Wang:2022jop, Abreu:2022sra, Wang:2022ztm}.

Because of the lack of experimental data or lattice QCD (LQCD) inputs, the unknown low-energy coupling constants (LECs) in ChEFT often have to be fixed using other models, as in Refs.~\cite{Du:2016tgp, Xu:2017tsr, Wang:2018atz, Meng:2019nzy, Wang:2019ato, Xu:2021vsi, Chen:2022iil, Xu:2025xrl, Liu:2025fhl}. In this work, however, we utilize the recently available LQCD data.
Recently, Xing \textit{et al.} performed the first LQCD study of $\Lambda_{c}\Lambda_{c}$ scattering using Lüscher's finite-volume method, extracting the scattering length and effective range at a pion mass $m_{\pi} \sim 303$ MeV and lattice spacing $a=0.07746$ fm~\cite{Xing:2025uai}. Their results indicate a repulsive interaction for the $I(J^P) = 0(0^+)$ $\Lambda_{c}\Lambda_{c}$ system. Coupled-channel effects involving $N\Xi_{cc}$ and $\Sigma_{c}\Sigma_{c}$ were also discussed. This LQCD calculation enables us to determine the corresponding LECs in the chiral Lagrangians by fitting to the lattice data, and then to extrapolate the results to the physical point.

This paper is organized as follows: In Sec.~\ref{Sec: potentials}, we introduce the chiral Lagrangians under flavor SU(2) and derive the effective baryon-baryon potentials. In Sec.~\ref{Sec: results}, we determine the contact-term LECs by fitting to the $S$-wave $\Lambda_{c}\Lambda_{c}$ phase shifts from LQCD. Using these fitted LECs, we then perform calculations at the physical pion mass, obtaining the $\Lambda_{c}\Lambda_{c}$ and $\Lambda_c \bar\Lambda_c$ scattering phase shifts and potentials, and we search for possible bound states by solving the Schr\"odinger equation. We also investigate the SU(3) effects induced by the two-kaon-exchange contribution and examine their impact on the binding properties of the $\Lambda_c\bar{\Lambda}_c$ systems. Finally, in Sec.~\ref{Sec: Summary}, we summarize our findings and discuss their implications for future studies of doubly heavy hexaquark systems.

\section{Chiral effective potentials}
\label{Sec: potentials}
\subsection{Effective Lagrangians}

In this work, we adopt the ChEFT Lagrangian in the heavy hadron formalism under the SU(2) case. We use Weinberg's power counting to study the  $\Lambda_{c}\Lambda_{c}(\bar\Lambda_{c})$ interactions up to the second chiral order $\mathcal{O}(\epsilon^2)$ at the one-loop level, and the order parameters are $\epsilon = q / \Lambda_{\chi}$, where $q$ represents the momentum of a Goldstone boson and parameter $\Lambda_{\chi}$ denotes the chiral symmetry broken scale or the mass of the heavy hadrons. 

In heavy hadron formalism, the lowest charmed baryons can be categorized into the following multiplets. 
The spin-$\frac{1}{2}$ isosinglet is defined as \cite{Chen:2022iil}

\begin{align}
	\psi_{1}=\begin{bmatrix}
		       0 & \Lambda_{c}^{+} \\
		       -\Lambda_{c}^{+} & 0 
	       \end{bmatrix} 
\quad {\mathrm{and}}\quad
	\psi^{c}_{1}=\begin{bmatrix}
		       0 & \bar\Lambda_{c}^{-} \\
		       -\bar\Lambda_{c}^{-} & 0 
	       \end{bmatrix} 
\end{align} 
for the antiparticle. The superscript $c$ represents charge conjugation.

The spin-$\frac{1}{2}$ and spin-$\frac{3}{2}$ isospin triplets are defined as \cite{Chen:2022iil,Wang:2019ato}

\begin{align}
	\psi_{3}=\begin{bmatrix}
	           \Sigma_{c}^{++}  & \frac{\Sigma_{c}^{+}}{\sqrt{2}} \\
	           \frac{\Sigma_{c}^{+}}{\sqrt{2}} & \Sigma_{c}^{0}  
	       \end{bmatrix}  , \quad
	\psi_{3^*}^{\mu}=\begin{bmatrix}
		       \Sigma_{c}^{*++}  & \frac{\Sigma_{c}^{*+}}{\sqrt{2}}\\
	           \frac{\Sigma_{c}^{*+}}{\sqrt{2}} & \Sigma_{c}^{*0}  
		      \end{bmatrix} ^{\mu}  ,
\end{align} 
and 
\begin{align}
	\psi^{c}_{3}=\begin{bmatrix}
	           \bar\Sigma_{c}^{--}  & \frac{\bar\Sigma_{c}^{-}}{\sqrt{2}} \\
	           \frac{\bar\Sigma_{c}^{-}}{\sqrt{2}} & \bar\Sigma_{c}^{0}  
	       \end{bmatrix}  , \quad
	\psi_{3^*}^{c\mu}=\begin{bmatrix}
		       \bar\Sigma_{c}^{*--}  & \frac{\bar\Sigma_{c}^{*-}}{\sqrt{2}}\\
	           \frac{\bar\Sigma_{c}^{*-}}{\sqrt{2}} & \bar\Sigma_{c}^{*0}  
		      \end{bmatrix} ^{\mu} 
\end{align} 
for the antiparticle.

The heavy baryon (anti) fields can be projected into the heavy components via the operators $\frac{1 \pm \slashed{v}}{2}$ \cite{Scherer:2012xha}:

\begin{align}
	\mathcal{B}_{i}=e^{i M_{i}v\cdot x}\frac{1+\slashed v}{2}\psi_{i},\qquad
	\mathcal{B}^{c}_{i}=e^{-i M_{i}v\cdot x}\psi^{c}_{i}\frac{1-\slashed{v}}{2},
\end{align}
where $\psi_{i}^{(c)}$ represents the heavy baryon fields $\psi_{1}^{(c)}$, $\psi_{3}^{(c)}$, or $\psi_{3^*}^{(c)\mu}$ and $M_{i}$ is the mass of the heavy baryon. $v$ is the four-velocity. 
We introduce the leading-order (LO) chiral Lagrangian \cite{Wang:2019ato, Chen:2022iil, Guo:2024pti}
\begin{align}
	\mathcal{L}_{\mathcal{B\phi }} =&~ \frac{1}{2} \mathrm{Tr} \big[\mathcal{\bar B}_{1}(iv \cdot D){\mathcal B}_{1} \big]+2 g_{2}\mathrm{Tr} \big[\mathcal{\bar B}_{3} \mathcal{S}\cdot u {\mathcal B}_{1}+\mathrm{H.c.} \big]\notag \\ 
    &+ g_{3}\mathrm{Tr} \big[ \mathcal{\bar B}^{\mu}_{3^*} u_{\mu } {\mathcal B}_{3}+\mathrm{H.c.} \big]
	+g_{4}\mathrm{Tr} \big[ \mathcal{\bar B}^{\mu}_{3^*}
	u_{\mu }  {\mathcal B}_{1}+\mathrm{H.c.} \big] \notag \\ 
    & +\mathrm{Tr} \big[\bar{\mathcal{B}}_{3}(iv\cdot D-\delta_{b}) \mathcal{B}_{3} \big] +2g_{5}\mathrm{Tr}\big[\mathcal{\bar B}^{\mu}_{3^*}
	\mathcal{S}\cdot u \mathcal{B}_{3^*\mu} \big]\notag \\ 
    & -\mathrm{Tr}\big[\mathcal{\bar B}^{\mu}_{3^*}(iv\cdot D-\delta_{c})\mathcal{ B}_{3^*\mu} \big]+2g_{1}\mathrm{Tr}[\mathcal{\bar B}_{3}
	\mathcal{S}\cdot u {\mathcal B}_{3} \big], \label{pi-La}
\end{align}
and  
\begin{align}
    \mathcal{L}_{\mathcal{B}^{c}\phi } =&~ \frac{1}{2} \mathrm{Tr} \big[\mathcal{\bar B}^{c}_{1} (iv \cdot D){\mathcal B}^{c}_{1} \big] -2 g_{2} \mathrm{Tr} \big[\mathcal{\bar B}^{c}_{3} \mathcal{S}\cdot u {\mathcal B}^{c}_{1}+\mathrm{H.c.} \big] \notag \\
    & - g_{3}\mathrm{Tr} \big[ \mathcal{\bar B}^{c\mu}_{3^*} u_{\mu } {\mathcal B}_{3}^{c} + \mathrm{H.c.} \big]
	-g_{4}\mathrm{Tr} \big[ \mathcal{\bar B}^{c\mu}_{3^*}
	u_{\mu }  {\mathcal B}^{c}_{1}+\mathrm{H.c.} \big] \notag \\ 
    & + \mathrm{Tr} \big[\bar{\mathcal{B}}^{c}_{3}(iv\cdot D-\delta_{b}) \mathcal{B}_{3}^{c} \big] -2g_{5}\mathrm{Tr} \big[ \mathcal{\bar B}^{c\mu}_{3^*} \mathcal{S}\cdot u \mathcal{B}^{c}_{3^*\mu} \big] \notag \\ 
    & - \mathrm{Tr}\big[\mathcal{\bar B}^{c\mu}_{3^*}(iv\cdot D-\delta_{c})\mathcal{ B}_{3^*\mu}^{c} \big] - 2g_{1}\mathrm{Tr}[\mathcal{\bar B}_{3}^{c} \mathcal{S}\cdot u {\mathcal B}_{3}^{c} \big], \label{pi-La 1}
\end{align}
for the pion-anticharmed baryon vertices, where Tr[$\cdots$] denotes the trace in flavor space and we have the spin operator $\mathcal{S}^{\mu}=(i/2) \gamma_{5} \sigma ^{\mu \nu } v_{\nu }$, the mass splittings $\delta_{b} =M_{\Sigma_{c}}-M_{\Lambda _{c}}$ and $\delta_{c} = M_{\Sigma^{*}_{c}}-M_{\Lambda_{c}}$, and the covariant derivative	
\begin{align}
	D_{\mu}\psi =\partial _{\mu}\psi+\Gamma_{\mu}\psi+\psi\Gamma_{\mu}^{T},
\end{align}
where $\Gamma_{\mu}^{T}$ is the transpose of $\Gamma_{\mu}$. 
The chiral connection $\Gamma_{\mu}$ and axial current $u_{\mu}$ are defined, respectively, as
\begin{align}
	\Gamma_{\mu}=\frac{1}{2} [\xi^{\dagger}  ,\partial _{\mu } \xi ],  \quad
	u_{\mu} =\frac{i}{2} \{ \xi^{\dagger}  ,\partial _{\mu } \xi\},
\end{align}
 with $\xi^{2} =U=\mathrm{exp} (\frac{i\phi }{f_{\pi} } )$, and
	\begin{align}
        \phi=~ &\begin{bmatrix}
			\pi^{0} & \sqrt{2} \pi^{+} \\
			\sqrt{2}  \pi^{-} & -\pi^{0}
		\end{bmatrix}.
	\end{align}
$\psi_{3}$ and $\psi_{3}^{*\mu}$ are degenerate in the heavy quark limit, which allows us to combine them into the superfields $\psi^{\mu}$ and $\bar{\psi}^{\mu}$ \cite{Wang:2019ato, Chen:2022iil}:
\begin{align}
	\psi^{\mu}=&\mathcal{B}_{3^{*}}-\frac{1}{\sqrt 3} (\gamma^{\mu}+v^{\mu})\gamma^{5}\mathcal{B}_{3},\\
	\bar{\psi}^{\mu}=&\mathcal{\bar B}_{3^{*}}+\frac{1}{\sqrt 3} \mathcal{\bar B}_{3}\gamma^{5}(\gamma^{\mu}+v^{\mu}).
\end{align} 

In terms of the superfields, the LO contact interaction Lagrangian for the charmed baryons takes a compact form \cite{Chen:2022iil}

\begin{widetext}
  
\begin{align}
	\mathcal{L} _{\mathcal{B B} }  = & C_{a}\mathrm{Tr} \big[\bar{\mathcal{B}} _{1} \mathcal{B}_{1} \big] \mathrm{Tr} \big[\bar{\mathcal{B}}_{1}\mathcal{B}_{1} \big]
	+ C_{b}\mathrm{Tr} \big[ \bar{\mathcal{B}}_{1} \mathcal{B}_{1} \big]
	\mathrm{Tr} \big[ \bar{\mathcal{\psi}}_{\mu} \mathcal{\psi}^{\mu} \big] + C_{c} \mathrm{Tr} \big[ \bar{\mathcal{\psi}}_{\mu}\mathcal{\psi}^{\mu} \big]
	\mathrm{Tr} \big[ \bar{\mathcal{\psi}}_{\nu}\mathcal{\psi}^{\nu} \big] + D_{a} \mathrm{Tr} \big[ \bar{\mathcal{\psi}}_{\mu}\tau ^{i} 
	\mathcal{\psi}^{\mu} \big] \mathrm{Tr} \big[\bar{\mathcal{\psi}}_{\nu}\tau _{i}\mathcal{\psi^{\nu}} \big] \notag \\ 
    & + i C_{d}\epsilon _{\mu \nu \alpha \beta} v^{\alpha} \mathrm{Tr} \big[ \bar{\mathcal{\psi}}_{\rho} \gamma ^{\beta} \gamma_{5} \tau ^{i} \mathcal{\psi^{\rho}} \big] \mathrm{Tr} \big[ \bar{\mathcal{{\psi }}}^{\mu}\mathcal{{\psi}}^{\nu} \big]
	+ i D_{b} \epsilon _{\mu \nu \alpha \beta} v^{\alpha} \mathrm{Tr} \big[ \bar{\mathcal{\psi}}_{\rho} \gamma ^{\beta} \gamma_{5} \tau_{i} \mathcal{\psi^{\rho}} \big] \mathrm{Tr} \big[ \bar{\mathcal{{\psi }}}^{\mu}\tau^{i}\mathcal{{\psi}}^{\nu} \big], \label{cont-La}
\end{align}
    
\end{widetext}  
where the $C_{a}$, $C_{b}$, $C_{c}$, $D_{a}$, $C_{d}$, and $D_{b}$ are six independent LECs and $\tau^i$ represents the Pauli matrix in isospin space. 
For the $\Lambda_{c}\bar{\Lambda}_{c}$ systems, the LO contact Lagrangian reads
\begin{align}
    \mathcal{L} _{\mathcal{B}_{1}^{c}\mathcal{B}_{1}}  = & C_{a}\mathrm{Tr} \big[\bar{\mathcal{B}}^{c} _{1} \mathcal{B}^{c}_{1} \big] \mathrm{Tr} \big[\bar{\mathcal{B}}_{1}\mathcal{B}_{1} \big] . \label{cont-La 1}
\end{align}

\begin{figure}[htbp]
   \centering
   \includegraphics[width=0.35\textwidth]{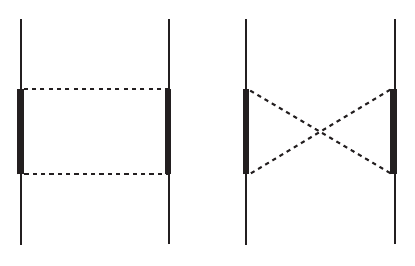}
   \captionsetup{justification=raggedright, singlelinecheck=false}
   \caption{Two-pion-exchange diagrams for the $\Lambda_{c}\Lambda_{c}$ system: the planar box diagram and the crossed box diagram. The thin, thick, and dashed lines denote the $\Lambda_{c}$, $\Sigma_{c}$, and $\pi$, respectively. } \label{loop}
\end{figure}

\subsection{Effective potentials}

Under the Breit approximation, the calculated scattering amplitudes $\mathcal{M} (\bm{q})$ constructed from the above Lagrangians can be related to the effective potential
\begin{align}
	\mathcal{V} (\bm{q} )=-\frac{\mathcal{M} (\bm{q})}{\sqrt{2M_{1}2M_{2}2M_{3}2M_{4}} }.
\end{align}
By expanding the Lagrangian~Eq. (\ref{cont-La}), we obtain the LO contact $\Lambda_{c}\Lambda_{c}$ potential
\begin{align}
	 \mathcal{V}^{\Lambda_{c}\Lambda_{c}}_{\rm Cont} =-8 C_{a}. \label{contp}
\end{align}

Now we turn to the next-to-leading order (NLO). At this order, there are two-pion-exchange (TPE) contributions; the diagrams are shown in Fig.~\ref{loop}. Note that the football and triangle diagrams vanish, since the $\Lambda_{c}\Lambda_{c}\pi\pi$ vertex does not exist. Because of the large mass splitting $\delta_{c}=M_{\Sigma^{*}_{c}}-M_{\Lambda_{c}} =233.5$ MeV, we do not consider the $\Sigma^{*}_{c}$ contributions explicitly; i.e., their effects are integrated out and absorbed into the contact LECs. Therefore, the expression for the TPE potential reads
\begin{align}
    \mathcal{V}^{\Lambda_{c}\Lambda_{c}}_{2\pi} = & \frac{3g_2^4}{4f_\pi^4} \Big[-\bm{q}^2J_{21}^B+\frac{2}{3}\bm{q}^2(\bm{\sigma}_1\cdot\bm{\sigma}_2)J_{21}^B +\bm{q}^4(J_{22}^B +2J_{32}^B \notag \\ 
    & +J_{43}^B) + 15 J_{41}^B -10\bm{q}^2(J_{31}^B+J_{42}^B) -\bm{q}^2 J_{21}^R \notag \\ 
    & - \frac{2}{3}\bm{q}^2 (\bm{\sigma}_1\cdot\bm{\sigma}_2)J_{21}^R +\bm{q}^4 (J_{22}^R+2J_{32}^R+J_{43}^R) \notag \\ 
    & + 15 J_{41}^R -10 \bm{q}^2 (J_{31}^R+J_{42}^R) \Big] (-\delta_{b},-\delta_{b}),
\label{2pep}
\end{align}  
where the definitions and the calculation procedures of the loop functions $J^{B/R}_{ij}$ are given in Refs.~\cite{Xu:2017tsr, Wang:2018atz, Xu:2021vsi, Xu:2025xrl, Liu:2025fhl}. 
The value of the $\bm{\sigma}_1\cdot\bm{\sigma}_2$ operator is calculated with the relation
\begin{align}
    \langle \bm{\sigma}_1\cdot\bm{\sigma}_2 \rangle = 2\big[ S(S+1) - S_{1}(S_{1}+1) - S_{2}(S_{2}+1) \big].\nonumber
\end{align}

Expanding the Lagrangian in Eq.~(\ref{cont-La 1}), we obtain the LO $\Lambda_{c}\bar \Lambda_{c}$ contact potential
\begin{align}
    \mathcal{V}^{\Lambda_{c}\bar \Lambda_{c}}_{\rm Cont} =4 C_{a} , \label{contp2}
\end{align}
where the $C_{a}$ is taken as in Eq.~\ref{contp}. The contribution from the annihilation term is ignored due to the lack of experimental data or LQCD calculations for $\Lambda_{c}\bar \Lambda_{c}$ scattering. 
The TPE potential for $\Lambda_{c}\bar{\Lambda}_{c}$ system is calculated as
\begin{align}
    \mathcal{V}^{\Lambda_{c}\bar \Lambda_{c}}_{2\pi} =\mathcal{V}^{\Lambda_{c} \Lambda_{c}}_{2\pi}. \label{tpe2}
\end{align}

In numerical calculations, a Gaussian regulator of the form $\mathcal{F}(\mathbf{q}) = \exp(-\mathbf{q}^{4} /\Lambda^{4})$ is introduced to regularize the ultraviolet divergence in the chiral potentials, where $\Lambda$ is the cutoff parameter \cite{Xu:2017tsr, Wang:2018atz, Xu:2021vsi, Entem:2017gor, Chen:2022iil}.

For the single-channel elastic scattering process, we calculate the two-body scattering amplitudes $T$ by solving the Lippmann-Schwinger equation,
\begin{align}
    T(p',p;\sqrt{s}) & =\mathcal{V}(p',p) +\int \frac{p''^{2}dp''}{(2\pi)^{3}}\mathcal{V}(p',p'') \notag \\ 
    & \quad\times \frac{2\mu}{p^{2}-p''^{2}+i\epsilon}T(p'',p;\sqrt{s}) ,\label{LSE}
\end{align}
with $\sqrt{s}$ the total energy of the two-body system  in the center-of-mass frame and $\mu$ the reduced mass. The transferred momentum $\bm{q}$ is related to the initial momentum $p=|\bm{p}|$ and the final momentum $p'=|\bm{p'}|$ in the center-of-mass frame via
\begin{align}
    \bm{q^2}=p^2+p'^2-2p p'\rm{cos}\theta, 
\end{align}
where $\theta$ is the scattering angle. Based on the effective potentials, we can solve the Lippmann-Schwinger equation in Eq. (\ref{LSE}) and then evaluate the $S$-wave $\Lambda_{c}\Lambda_{c}(\bar \Lambda_{c})$ scattering phase shift using
\begin{align}
    {\rm {tan}} \delta_{0}=\frac{{\rm Im}(T_{0})}{{\rm Re}(T_{0})}. 
\end{align}

To understand the potential behaviors of the $\Lambda_{c} \Lambda_{c}$ and $\Lambda_{c} \bar \Lambda_{c}$ interactions, we need to make Fourier transformation on $\mathcal{V}(\mathbf{q})$ to get $\mathcal{V}(r)$, the potentials in coordinate space:
\begin{align}
     \mathcal{V}(r)=\int\frac{d^3\mathbf{q}}{(2\pi)^3}\mathcal{V}(\mathbf{q})e^{-i\mathbf{q}\cdot\mathbf{r}}\mathcal{F}(\mathbf{q}).
\end{align}
Furthermore, based on $\mathcal{V}(r)$, we can solve the Schrödinger equations to investigate whether the $\Lambda_{c} \Lambda_{c}$ and $\Lambda_{c} \bar \Lambda_{c}$ systems can form bound states. 

\section{Numerical results and discussion}
\label{Sec: results}
\subsection{$\Lambda_{c}\Lambda_{c}$ interaction }\label{A}

The constant $C_{a}$ in the contact potential~\ref{contp} shall be determined by fitting to the LQCD data~\cite{Xing:2025uai}. For the first time in Ref.~\cite{Xing:2025uai}, a LQCD calculation was performed for the $I(J^P)=0(0^+)$ $\Lambda_{c}\Lambda_{c}$ scattering on two $N_{f}=2+1$ Wilson-Clover ensembles with the pion mass $m_{\pi}\sim303$ MeV and lattice spacing $a=0.07746$ fm. They obtained the scattering phase shift and determined the scattering length. 

In Table~\ref{mass}, we list the corresponding pion and charmed baryon masses given in Ref.~\cite{Xing:2025uai}, which will be used in the fitting procedure later. For $g_2$ in Eq.~(\ref{2pep}), we adopt $g_{2}=-0.6$ determined from the decay widths of the process $\Sigma_{c}\to\Lambda_{c}\pi$ \cite{Chen:2022iil, ParticleDataGroup:2020ssz}. Note that $C_a$ and $g_{2}$ are assumed to be pion mass independent \cite{Haidenbauer:2017dua, Song:2020isu}. The dependence of the pion decay constant $f_{\pi}$ on the pion mass has been studied in Ref.~\cite{BMW:2013fzj}, and we adopt $f_{\pi}\sim93$ MeV for the physical pion mass $m_{\pi}\sim138$ MeV and $f_{\pi}\sim101$ MeV for $m_{\pi}\sim303$ MeV. We have checked that neglecting the pion mass dependence of $f_{\pi}$ does not affect our conclusions. 

\begin{figure}[ht]
   \centering
   \includegraphics[width=0.48\textwidth]{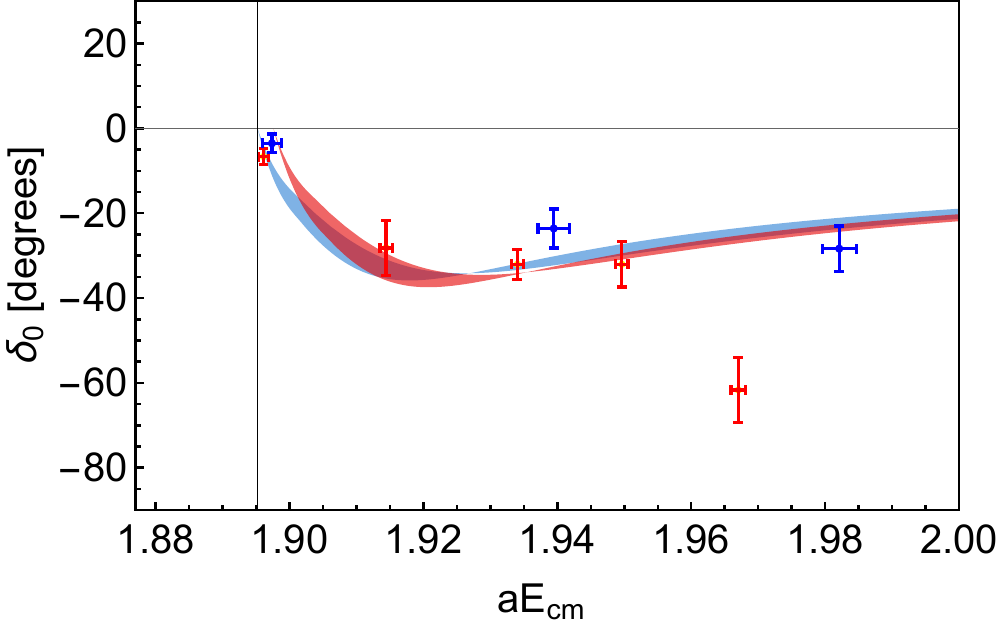}
   \captionsetup{justification=raggedright, singlelinecheck=false}
   \caption{$0(0^+)$ $\Lambda_c\Lambda_c$ scattering phase shifts as a function of $aE_{\rm cm}$. The blue and red dots denote the LQCD data from Ref. \cite{Xing:2025uai} for the F32P30 and F48P30 ensembles, respectively. The gray-blue and light-red bands correspond to our ChEFT fits of the F32P30 and F48P30 ensembles, respectively, obtained by varying the cutoff $\Lambda$ from 0.56 to 0.66 GeV. } \label{psfit}
\end{figure}

The data points of the LQCD simulations, as well as our fitted results of the $0(0^+)$ $\Lambda_c\Lambda_c$ scattering phase shifts, are presented in Fig.~\ref{psfit}, where the F32P30 and F48P30 ensembles are both considered. Note that we exclude the data points at high energy from each ensemble as in Ref.~\cite{Xing:2025uai}. The parameters $C_a$ with uncertainties are determined by minimizing the $\chi^2$/d.o.f.. It is observed that the fitted $C_a$ gradually decreases as the cutoff $\Lambda$ increases. 
\renewcommand{\arraystretch}{1.3}
\begin{center}
  \begin{table}[hb]
  	\centering
    \captionsetup{justification=raggedright, singlelinecheck=false}
  	\caption{Baryon masses for different pion masses adopted in the LQCD simulation in Ref.~\cite{Xing:2025uai}.}\label{mass}
  	\setlength{\tabcolsep}{3.6mm}{
  	\begin{tabular}{ccccc}
  	  \toprule[1.5pt]
        \toprule[1pt]
  		Ensemble &  {$m_{\pi}$/MeV} &  {$M_{\Lambda_{c}}$/GeV} &  {$M_{\Sigma_{c}}$/GeV}   \\
  		\midrule[0.7pt]
  		F32P30  & 303.2   & 2.413   &  2.572 \\
  		F48P30  & 303.4   & 2.410   &  2.566 \\
  		\bottomrule[1pt]
        \bottomrule[1.5pt]
  	\end{tabular}}
  \end{table}
\end{center}

Here, we give a comment about the finite-volume effects of the lattice data we used. Although the Lüscher quantization condition converts the leading power-law volume dependence of the two-particle energy levels into the infinite-volume phase shifts, residual corrections of the order of $e^{-m_\pi L}$ remain. The two ensembles used in Ref.~\cite{Xing:2025uai}, $m_\pi L=3.81$ and $5.72$, correspond to exponential suppression factors of about $2.2\%$ and $0.33\%$, respectively. Therefore, such residual finite-volume effects are expected to be small for the present analysis.

\begin{figure}[htbp]
   \centering
   \includegraphics[width=0.48\textwidth]{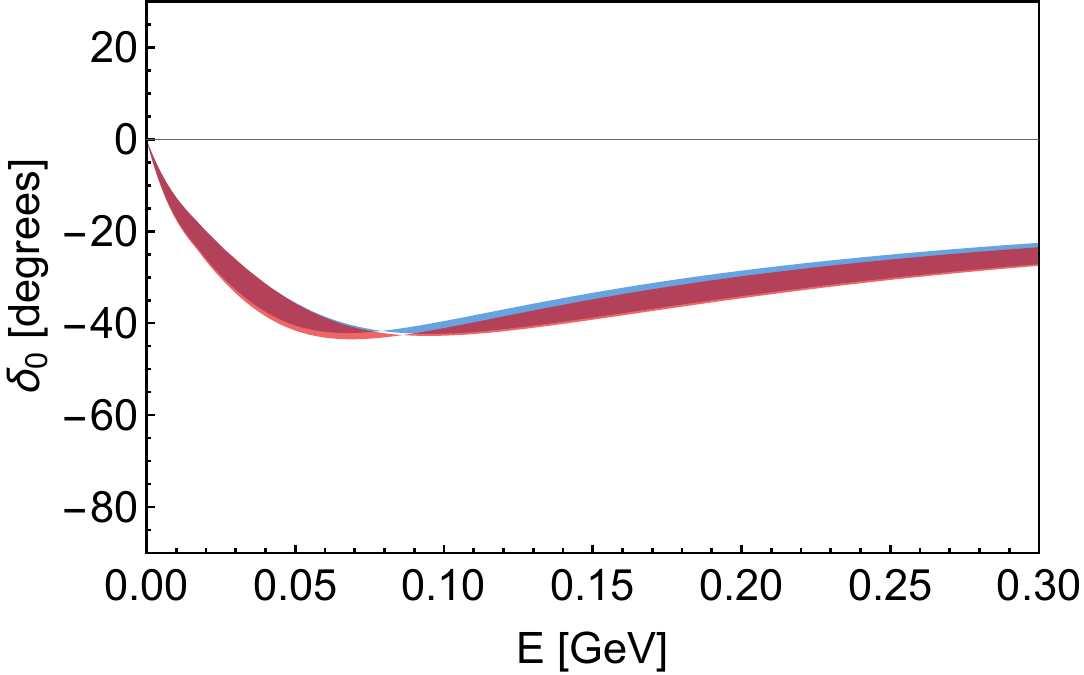}
   \captionsetup{justification=raggedright, singlelinecheck=false}
   \caption{$0(0^+)$ $\Lambda_{c}\Lambda_{c}$ scattering phase shifts as a function of the kinetic energy in the center-of-mass frame at the physical pion mass. The gray-blue and light-red bands correspond to our ChEFT fits of the F32P30 and F48P30 ensembles, respectively, obtained by varying the cutoff $\Lambda$ from 0.56 to 0.66 GeV. } \label{ps}
\end{figure}

Then, the fitted $C_{a}$ can be used to estimate the $\Lambda_{c}\Lambda_{c}$ scattering phase shifts at the physical pion mass. We therefore calculate the $0(0^+)$ $\Lambda_{c}\Lambda_{c}$ scattering phase shifts for $m_{\pi}=138$ MeV using two sets of $C_{a}$, which are presented in Fig.~\ref{ps}. 

\renewcommand{\arraystretch}{1.2}  
\begin{table*}[htbp]
    \centering
    \captionsetup{justification=raggedright, singlelinecheck=false}
    \caption{Fitted values of constant $C_{a}$ with uncertainty (in units of $\mathrm{GeV}^{-2}$) in the $\Lambda_{c}\Lambda_{c}$ contact potential and the corresponding $\chi^{2}/\mathrm{d.o.f.}$, obtained using the F32P30 and F48P30 ensembles for various cutoff values $\Lambda$  (in units of $\mathrm{GeV}$).}\label{fit}
    \setlength{\tabcolsep}{0.8pt}  
    \begin{tabular*}{\textwidth}{@{\extracolsep{\fill}}ccccccc@{}}
        \toprule[1.5pt]
        \toprule[1pt]
        $\Lambda$ & 0.56 & 0.58 & 0.60 & 0.62 & 0.64 & 0.66 \\
        \midrule[0.7pt]
        $C_a$ (F32P30) 
        & $-4.04 \pm 0.19$ & $-3.75 \pm 0.18$ & $-3.71 \pm 0.17$ & $-3.60 \pm 0.16$ & $-3.46 \pm 0.16$ & $-3.38 \pm 0.15$ \\
        $\chi^{2}/\mathrm{d.o.f.}$ 
        & 6.40 & 6.33 & 6.15 & 6.04 & 5.94 & 5.84 \\
        $C_a$ (F48P30) 
        & $-3.90 \pm 0.19$ & $-3.73 \pm 0.18$ & $-3.65 \pm 0.17$ & $-3.55 \pm 0.16$ & $-3.44 \pm 0.16$ & $-3.36 \pm 0.15$ \\
        $\chi^{2}/\mathrm{d.o.f.}$ 
        & 6.39 & 5.92 & 5.47 & 5.11 & 4.81 & 4.56 \\
        \bottomrule[1pt]
        \bottomrule[1.5pt]
    \end{tabular*}
\end{table*}

The phase shifts obtained at the physical pion mass show only a weak dependence on $C_{a}$ and cutoff. Our calculation confirms that the $0(0^+)$ $\Lambda_{c}\Lambda_{c}$ interaction remains repulsive, consistent with the LQCD studies \cite{Xing:2025uai}. 

\begin{figure}[htbp]
    \centering
        \includegraphics[width=\linewidth]{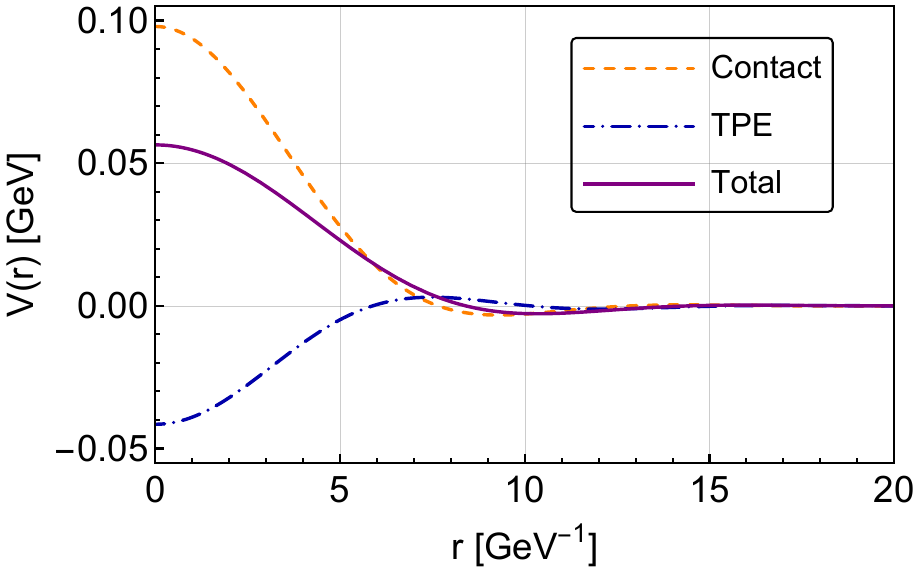}  
        \label{v1}
    \captionsetup{justification=raggedright, singlelinecheck=false}
    \caption{$0(0^{+})$ $\Lambda_{c}\Lambda_{c}$ effective potentials in coordinate space at the physical pion mass, with cutoff $\Lambda = 0.60$ GeV. Here we set $C_{a} = -3.65~\mathrm{GeV}^{-2}$ from Table \ref{fit}.}
    \label{potential}
\end{figure}

To illustrate the characteristics of this interaction, we present in Fig.~\ref{potential} the $0(0^+)$ $\Lambda_{c}\Lambda_{c}$ effective potentials in coordinate space, computed with  $\Lambda = 0.60$ GeV at the physical pion mass. The calculation shows that the attractive TPE contribution is considerably weaker than the repulsive contact term, resulting in an overall repulsive potential. It contrasts the results obtained with those in Refs.~\cite{Lee:2011rka, Liu:2012zzo} but is consistent with the results in Refs.~\cite{Huang:2013rla, Chen:2022iil}. 

Finally, we discuss the $\chi^2/\mathrm{d.o.f.}$ obtained in Table \ref{fit}. It should be noted that only a limited number of phase shift data points are available in the fitting region, resulting in a small number of degrees of freedom. Consequently, the obtained $\chi^2/\mathrm{d.o.f.}$ values are relatively large. The limited number of the data points can indeed affect the precision of the fitted parameters and the obtained interaction potentials. However, we believe the general behaviors of our results are not affected.

\subsection{$\Lambda_{c}\bar \Lambda_{c}$ interaction }\label{B}

\begin{figure*}[ht]
    \centering
    \begin{subfigure}{0.46\textwidth}
        \includegraphics[width=\linewidth]{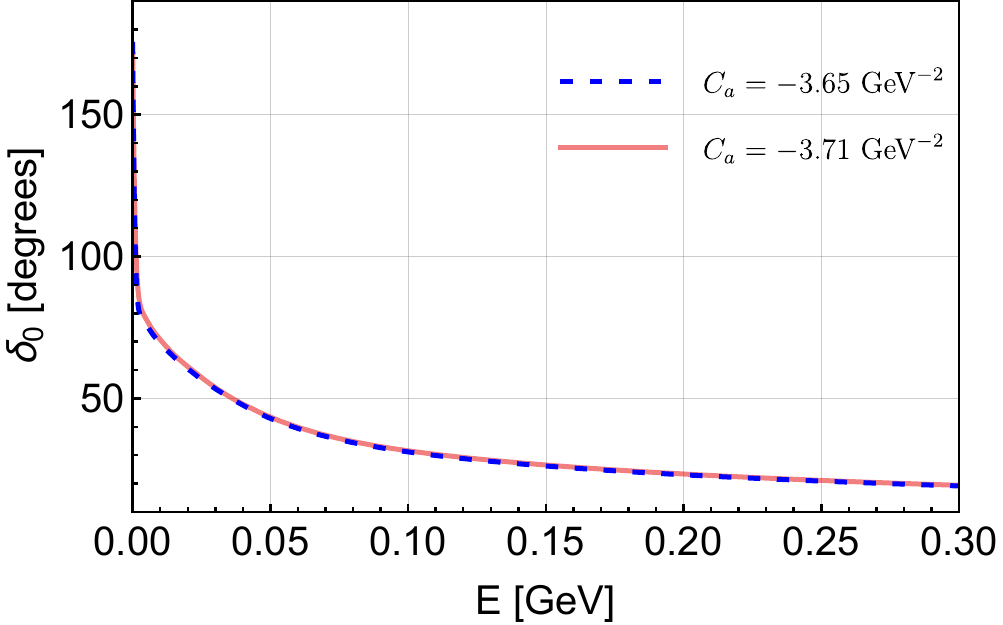}
        \caption{$0(0^{-+})$ $\Lambda_{c}\bar \Lambda_{c}$}
        \label{ps0}
    \end{subfigure}
    \begin{subfigure}{0.46\textwidth}
        \includegraphics[width=\linewidth]{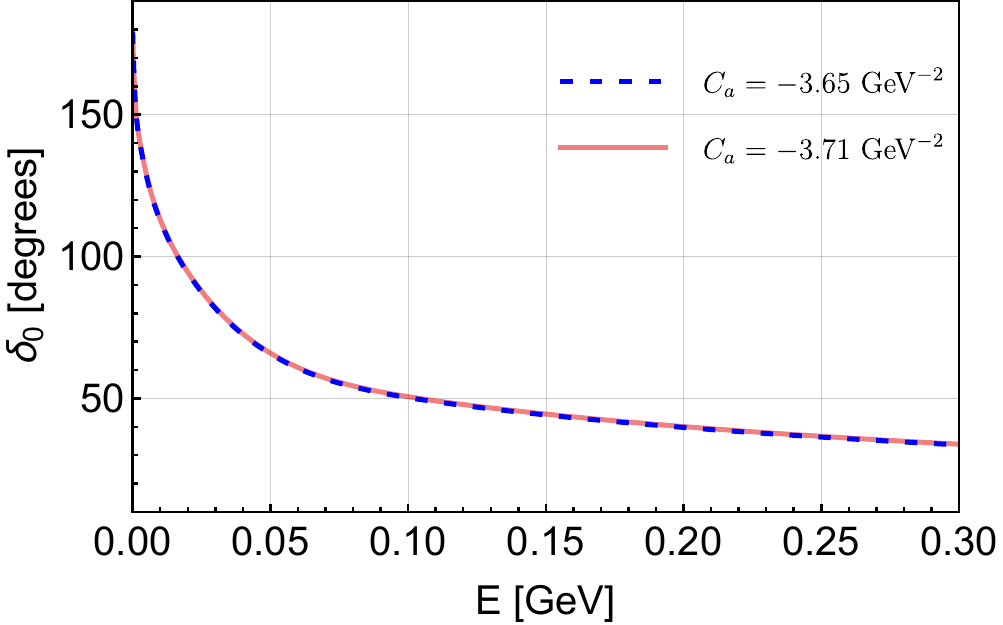} 
        \caption{$0(1^{--})$ $\Lambda_{c}\bar \Lambda_{c}$}
        \label{ps1}
    \end{subfigure}
    \captionsetup{justification=raggedright, singlelinecheck=false}
    \caption{$0(0^{-+})$ and $0(1^{--})$ $\Lambda_{c}\bar{\Lambda}_{c}$ scattering phase shifts as a function of the kinetic energy $E$ in the center-of-mass frame at the physical pion mass, with cutoff $\Lambda = 0.60$ GeV. }
    \label{ps0-1}
\end{figure*}

\begin{figure*}[ht]
    \centering
    \begin{subfigure}{0.46\textwidth}
        \centering
        \includegraphics[width=\linewidth]{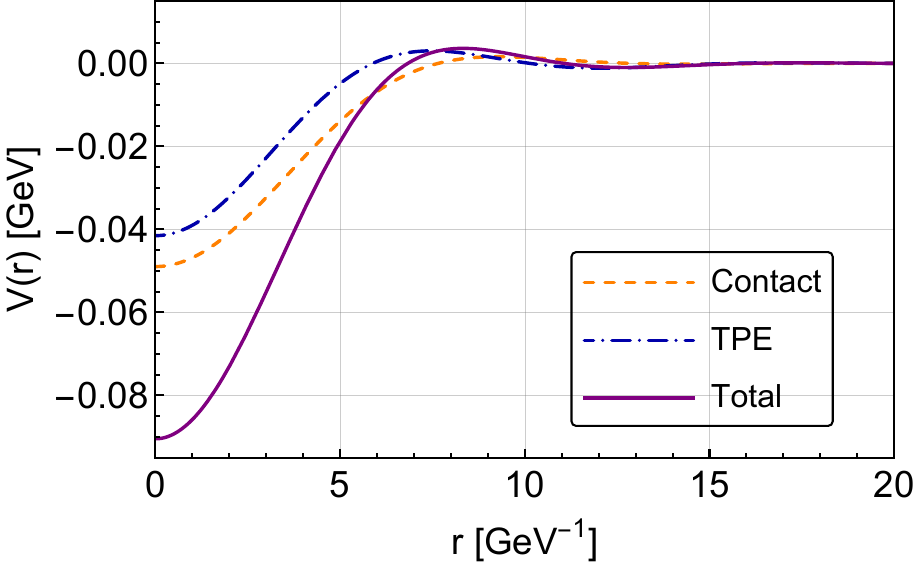}
        \caption{$0(0^{-+})$ $\Lambda_{c}\bar \Lambda_{c}$}
        \label{v0}
    \end{subfigure}
    \begin{subfigure}{0.46\textwidth}
        \centering
        \includegraphics[width=\linewidth]{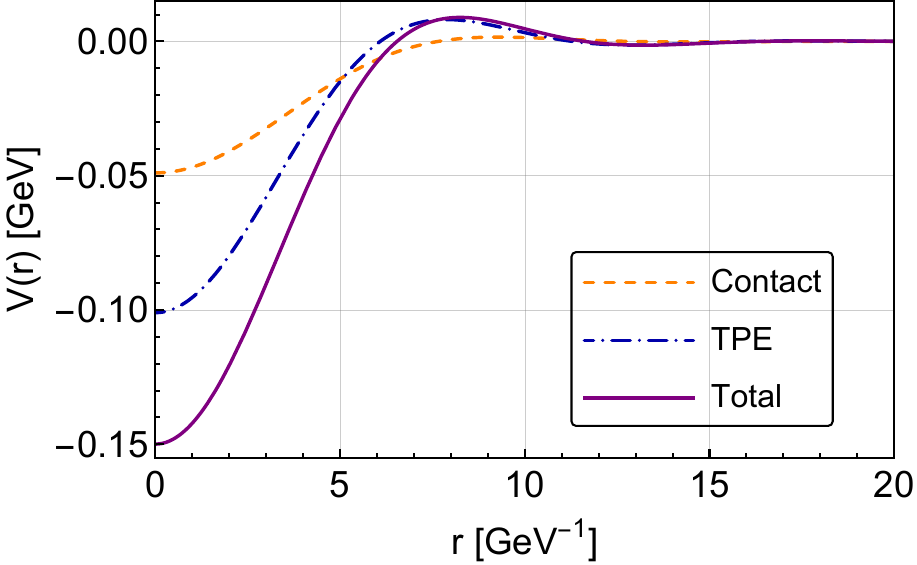} 
        \caption{$0(1^{--})$ $\Lambda_{c}\bar \Lambda_{c}$}
        \label{v1}
    \end{subfigure}
    \captionsetup{justification=raggedright, singlelinecheck=false}
    \caption{The $0(0^{-+})$ and $0(1^{--})$ $\Lambda_{c}\bar{\Lambda}_{c}$ effective potentials in coordinate space at the physical pion mass using the cutoff $\Lambda = 0.60$ GeV and the fitted $C_{a} = -3.65~\mathrm{GeV}^{-2}$ in Table \ref{fit}.}
    \label{v0-1}
\end{figure*}

Employing the effective potentials given in Eqs.~(\ref{contp2}) and (\ref{tpe2}), we calculate the $0(0^{-+})$ and $0(1^{--})$ $\Lambda_{c}\bar{\Lambda}_{c}$ scattering phase shifts as functions of the center-of-mass kinetic energy at the physical pion mass, as well as the effective potentials in coordinate space, using a fixed cutoff $\Lambda = 0.60$ GeV. The results are presented in Figs.~\ref{ps0-1} and \ref{v0-1}, respectively.

As illustrated in Fig.~\ref{ps0}, the phase shift of the $0(0^{-+})$ channel begins at $180^\circ$ and decreases as the center-of-mass kinetic energy increases, indicating an attractive interaction. According to Levinson's theorem \cite{Taylor:1972pty}, this behavior suggests the existence of a bound state. Also, we find that the two curves nearly coincide due to the similar $C_a$ values in the two sets. In the $0(1^{--})$ channel [Fig.~\ref{ps1}], the phase shift also decreases with increasing center-of-mass kinetic energy, so this channel possesses an attractive interaction. The consistency of the two channels originates from their common contact potentials [Eq.~(\ref{contp2})].
\renewcommand{\thesubfigure}{(\alph{subfigure})}

\renewcommand{\arraystretch}{1.6} 
\begin{table}[htbp]
    \centering
    \captionsetup{justification=raggedright, singlelinecheck=false}
    \caption{Dependence of the binding energies (in units of MeV) on the cutoff $\Lambda$ (in units of MeV) for the $0(0^{-+})$ and $0(1^{--})$ $\Lambda_{c}\bar{\Lambda}_{c}$ systems. The results are obtained using the fitted $C_{a}$ values from F32P30 and F48P30 ensembles, respectively (see Table \ref{fit}).}\label{be}
    \setlength{\tabcolsep}{10pt}
    \begin{tabular}{ccccc}
        \toprule[1.5pt]
        \toprule[1pt]
        $\Lambda$ & \multicolumn{2}{c}{$0(0^{-+})$} & \multicolumn{2}{c}{$0(1^{--})$} \\
        \cmidrule{2-3}\cmidrule{4-5}
        & F32P30 & F48P30 & F32P30 & F48P30 \\
        \midrule[0.7pt]
        560 & $0.2^{+0.2}_{-0.1}$ & $0.1^{+0.2}_{-0.1}$ & $8.6^{+0.8}_{-0.7}$ & $8.1^{+0.7}_{-0.8}$ \\
        580 & $0.3^{+0.2}_{-0.2}$ & $0.2^{+0.2}_{-0.1}$ & $10.7^{+0.8}_{-0.9}$ & $10.5^{+0.9}_{-0.8}$ \\
        600 & $0.5^{+0.4}_{-0.1}$ & $0.5^{+0.3}_{-0.2}$ & $14.5^{+0.9}_{-1.0}$ & $14.1^{+1.0}_{-0.9}$ \\
        620 & $1.0^{+0.4}_{-0.3}$ & $0.9^{+0.4}_{-0.3}$ & $18.6^{+1.1}_{-1.0}$ & $18.3^{+1.1}_{-1.0}$ \\
        640 & $1.5^{+0.5}_{-0.4}$ & $1.5^{+0.4}_{-0.5}$ & $23.5^{+1.2}_{-1.3}$ & $23.3^{+1.2}_{-1.2}$ \\
        660 & $2.5^{+0.5}_{-0.6}$ & $2.3^{+0.6}_{-0.5}$ & $29.7^{+1.3}_{-1.3}$ & $29.5^{+1.3}_{-1.3}$ \\
        \bottomrule[1pt]
        \bottomrule[1.5pt]
    \end{tabular}
\end{table}
Then, we present the effective potentials in coordinate space obtained with the fitted $C_a$ at $\Lambda = 0.60$ GeV in Fig.~\ref{v0-1}. For both the $0(0^{-+})$ and $0(1^{--})$ channels in Fig.~\ref{v0-1}, the LO contact interactions and NLO TPE interactions are all attractive. Compared to the $0(0^{-+})$, the attraction of the $0(1^{--})$ TPE potential is stronger, suggesting that the $0(1^{--})$ channel may support a more deeply bound state. Now we proceed to solve the Schrödinger equation to locate the $0(0^{-+})$ and $0(1^{--})$ $\Lambda_{c}\bar{\Lambda}_{c}$ bound states. After the calculations, the dependences of the binding energies on the cutoff $\Lambda$ are listed in Table~\ref{be}. At $\Lambda = 0.56$ GeV, a very shallow bound state starts to appear in the $0(0^{-+})$ channel, while a deeper bound state is found in the $0(1^{--})$ channel. As $\Lambda$ increases, the binding energies of both states increase accordingly.

It is worth mentioning the mass splitting between the $0(0^{-+})$ and $0(1^{--})$ channels. In Refs.~\cite{Lee:2011rka, Song:2022yfr}, the $0(0^{-+})$ and $0(1^{--})$ $\Lambda_{c}\bar{\Lambda}_{c}$ systems produced the bound states with almost the same binding energy. It is caused by the fact that their spin-dependent terms were negligible in the $S$-wave $\Lambda_{c}\bar{\Lambda}_{c}$ interaction. In our work, however, the spin-spin terms originated from the TPE potentials play a decisive role, leading to a distinct mass splitting between the $0(0^{-+})$ and $0(1^{--})$ channels. 

So, the splitting caused by the TPE can be treated as a key feature in our ChEFT framework. This obvious discrepancy between ChEFT and the model calculations~\cite{Lee:2011rka, Song:2022yfr} can serve as a benchmark in future lattice and experimental verifications. 

Here, we can give some comments about the $\Lambda_{c}\bar{\Lambda}_{c}$ annihilation contribution. The annihilation terms are usually treated as short-distance dynamics. For example, the annihilation contribution was parametrized, and the corresponding parameters are determined from antibaryon-baryon scattering phase shifts \cite{Kang:2013uia, Dai:2017ont, Xiao:2024jmu}. In our work, due to the lack of direct constraints from $\Lambda_c\bar{\Lambda}_c$ scattering or annihilation observables, the parametrized annihilation terms cannot be determined in the present work. However, Ref.~\cite{Guo:2024pti} studied the experimental $e^+e^- \rightarrow \Lambda_c\bar{\Lambda}_c$ cross section data, parametrized, and constrained the annihilation contributions. The results showed that the annihilation terms are not the main contribution near threshold.

Regarding experimental candidates, the $Y(4630)$ reported by the Belle Collaboration is unlikely to be a $\Lambda_c \bar{\Lambda}_c$ molecule, as its mass peak lies significantly above the $\Lambda_c \bar{\Lambda}_c$ threshold \cite{Belle:2008xmh, Song:2022yfr}. In Ref.~\cite{Dong:2021juy}, a $\Lambda_{c}\bar{\Lambda}_{c}$ molecule was identified, with its pole position determined by fitting to the BESIII data~\cite{BESIII:2017kqg}. Subsequently, an analysis of the $e^+e^- \to \Lambda_c \bar{\Lambda}_c$ data~\cite{BESIII:2023rwv} predicted a $\Lambda_{c} \bar{\Lambda}_{c}$ bound state $\sim38$ MeV below the threshold using a single-channel approach~\cite{Salnikov:2023qnn}, while a coupled-channel study in Ref.~\cite{Nakamura:2023obk} found a similar pole but located above the threshold. Our results are consistent with these studies that directly analyze the experimental data.
Based on our calculations, we suggest searching for the $0(0^{-+})$ $\Lambda_c \bar{\Lambda}_c$ molecular state in the $D \bar{D}^*$ and $D^* \bar{D}^*$ final states. Meanwhile, the $0(1^{--})$ state should be sought in the $D \bar{D}$, $D \bar{D}^*$, and $D^* \bar{D}^*$ channels. Such heavy baryonium states may be accessible in, e.g., the $B$-meson decays, $e^+ e^-$ and $p\bar{p}$ collisions at the LHCb, BESIII, Belle II, and the future $\bar{\text{P}}$ANDA experiment. 

\subsection{The estimation of the SU(3) effects}\label{C} 

\begin{figure}[htbp]
   \centering
   \includegraphics[width=0.18\textwidth]{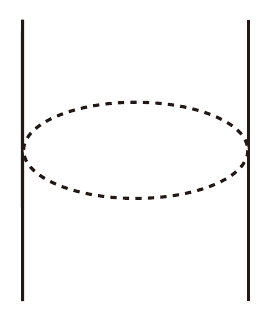}
   \captionsetup{justification=raggedright, singlelinecheck=false}
   \caption{Two-kaon-exchange football diagram for the $\Lambda_{c}\Lambda_{c}$ system. The thin and dashed lines denote the $\Lambda_{c}$ and $K$, respectively. } \label{loop2}
\end{figure}

To further investigate the flavor effects beyond the SU(2) case, we extend the interaction to the SU(3) symmetry framework by including the strange pseudoscalar mesons. In the SU(3) case, the spin-$\frac{1}{2}$ and spin-$\frac{3}{2}$ charmed baryon fields can be organized as antisymmetric and symmetric $3\times3$ matrices, respectively. Meanwhile, the $\eta$ and $K$ mesons are included in the pseudoscalar meson field $\phi$. Since the concerned vertices are negligible, we do not consider the $\eta$ exchange here \cite{Guo:2024pti}. 
Therefore, the dominant SU(3) correction comes from the two-kaon-exchange (TKE) term at NLO, as illustrated by the football diagram in Fig.~\ref{loop2}. The corresponding potential is given by
\begin{align}
    \mathcal{V}^{\Lambda_{c}\Lambda_{c}}_{2K} = \frac{1}{2f_\pi^4}J^F_{22}.
\label{2kep}
\end{align}

Including the TKE term, we can obtain the total potential
\begin{align}
    \mathcal{V}^{\Lambda_{c}\Lambda_{c}}_{\rm Total}
    &=
    \mathcal{V}^{\Lambda_{c}\Lambda_{c}}_{\rm Cont}
    +
    \mathcal{V}^{\Lambda_{c}\Lambda_{c}}_{2\pi}
    +
    \mathcal{V}^{\Lambda_{c}\Lambda_{c}}_{2K}.
\label{toltal}
\end{align}

Then, we can refit the LQCD results and present the resulting $C_a$ with uncertainties in Table~\ref{fit2}. $m_{K}\sim524$ MeV is adopted for $m_{\pi}\sim303$ MeV \cite{Xing:2025uai}. Compared with the fitted values in the SU(2) case, the values of $C_a$ shift after including the SU(3) effects. 
\renewcommand{\arraystretch}{1.2}  
\begin{table*}[htbp]
    \centering
    \captionsetup{justification=raggedright, singlelinecheck=false}
    \caption{Refitted values of $C_{a}$ with uncertainty (in units of $\mathrm{GeV}^{-2}$) and the corresponding $\chi^{2}/\mathrm{d.o.f.}$ for various cutoff values $\Lambda$ (in units of $\mathrm{GeV}$) with SU(3) effects.}\label{fit2}
    \setlength{\tabcolsep}{0pt}  
    \begin{tabular*}{\textwidth}{@{\extracolsep{\fill}}ccccccc@{}}
        \toprule[1.5pt]
        \toprule[1pt]
        $\Lambda$ & 0.56 & 0.58 & 0.60 & 0.62 & 0.64 & 0.66 \\
        \midrule[0.7pt]
        $C_a$ (F32P30) 
        & $-4.98 \pm 0.19$ & $-4.83 \pm 0.18$ & $-4.69 \pm 0.17$ & $-4.57 \pm 0.16$ & $-4.46 \pm 0.16$ & $-4.37 \pm 0.15$ \\
        $\chi^{2}/\mathrm{d.o.f.}$  
        & 6.41 & 6.28 & 6.16 & 6.06 & 5.95 & 5.86 \\
        $C_a$ (F48P30) 
        & $-4.85 \pm 0.19$ & $-4.72 \pm 0.18$ & $-4.61 \pm 0.17$ & $-4.51 \pm 0.16$ & $-4.42 \pm 0.16$ & $-4.33 \pm 0.15$ \\
        $\chi^{2}/\mathrm{d.o.f.}$  
        & 6.50 & 6.00 & 5.57 & 5.20 & 4.90 & 4.64 \\
        \bottomrule[1pt]
        \bottomrule[1.5pt]
    \end{tabular*}
\end{table*}

For the $\Lambda_c\bar{\Lambda}_c$ interaction, the TKE contribution has an opposite sign compared with that in the $\Lambda_c\Lambda_c$ channel
\begin{align}
    \mathcal{V}^{\Lambda_{c}\bar \Lambda_{c}}_{2K}
    = -\mathcal{V}^{\Lambda_{c}\Lambda_{c}}_{2K}.
\label{2kep2}
\end{align}

Using the $C_a$ in Table~\ref{fit2}, we solve the Schrödinger equation to search for the $0(0^{-+})$ and $0(1^{--})$ $\Lambda_c\bar{\Lambda}_c$ bound states at physical pion and kaon masses. The cutoff dependences of the binding energies are shown in Table~\ref{be2}.
\renewcommand{\arraystretch}{1.65} 
\begin{table}[htbp]
    \centering
    \captionsetup{justification=raggedright, singlelinecheck=false}
    \caption{Dependence of the binding energies (in units of MeV) on the cutoff $\Lambda$ (in units of MeV) for the $0(0^{-+})$ and $0(1^{--})$ $\Lambda_{c}\bar{\Lambda}_{c}$ systems. The results are obtained using the refitted $C_{a}$ in Table \ref{fit2}. ``$\cdots$" means that there is no bound state.}\label{be2}
    \setlength{\tabcolsep}{17pt}  
    \begin{tabular}{cccc}
        \toprule[1.5pt]
        \toprule[1pt]
        $\Lambda$ & $0(0^{-+})$ & \multicolumn{2}{c}{$0(1^{--})$} \\
        \cmidrule{3-4}
        & & F32P30 & F48P30 \\
        \midrule[0.7pt]
        560 & $\cdots$ & $3.4^{+0.6}_{-0.6}$ & $3.0^{+0.6}_{-0.6}$ \\
        580 & $\cdots$ & $4.9^{+0.7}_{-0.6}$ & $4.5^{+0.6}_{-0.7}$ \\
        600 & $\cdots$ & $7.0^{+0.8}_{-0.8}$ & $6.6^{+0.8}_{-0.7}$ \\
        620 & $\cdots$ & $9.7^{+1.0}_{-0.8}$ & $9.4^{+0.9}_{-0.9}$ \\
        640 & $\cdots$ & $13.2^{+1.0}_{-1.1}$ & $13.0^{+1.0}_{-1.1}$ \\
        660 & $\cdots$ & $17.7^{+1.2}_{-1.2}$ & $17.4^{+1.1}_{-1.2}$ \\
        \bottomrule[1pt]
        \bottomrule[1.5pt]
    \end{tabular}
\end{table}

Compared with the SU(2) case, the interactions here in the two channels are still attractive, and the discrepancy among them emphasized in Sec.~\ref{B} still exists. Note that the inclusion of $KK$ exchange weakens the attraction in both channels, causing the shallow bound state in the $0(0^{-+})$ channel to disappear and reducing the binding energy in the $0(1^{--})$ channel. In a word, in the SU(3) case, although both channels are still attractive, only the $0(1^{--})$ bound state remains.

\section{summary}
\label{Sec: Summary}

Employing a ChEFT framework with a common set of LECs constrained by LQCD simulations, we simultaneously investigate the $S$-wave interactions of the $\Lambda_c\Lambda_c$ and $\Lambda_c \bar{\Lambda}_c$ systems. Specifically, we calculate the $\Lambda_c\Lambda_c$ and $\Lambda_c\bar{\Lambda}_c$ potentials up to NLO, incorporating both contact and TPE contributions. The common LECs are determined by fitting to the LQCD results for the $I(J^P) = 0(0^+)$ $\Lambda_c\Lambda_c$ phase shifts at a pion mass of approximately $303$ MeV. Using these LECs, we then explore the interaction properties at the physical pion mass, including phase shifts, coordinate space potentials, and the possible formation of bound states. The robustness of this study lies in its direct connection to LQCD input.

In the $\Lambda_c\Lambda_c$ sector, our results consistently support a repulsive nature for the $0(0^+)$ channel. A detailed analysis of the effective potential in coordinate space reveals that the repulsive contact term dominates over the attractive two-pion-exchange contribution, yielding an overall repulsive potential.

For the $\Lambda_c\bar{\Lambda}_c$ system, the phase shifts clearly indicate attractive interactions. Further analysis of the coordinate space effective potential and a search for bound state solutions show that the attraction is more pronounced in the $0(1^{--})$ channel, where a bound state is more likely to exist. The spin-spin interaction term, which is non-negligible in our effective potential, emerges as the key factor distinguishing the dynamics of the $0(0^{-+})$ and $0(1^{--})$ $\Lambda_c \bar\Lambda_c$ systems-a refinement not addressed in earlier works~\cite{Lee:2011rka, Song:2022yfr}.

In summary, our results reveal that the $\Lambda_c\Lambda_c$ channel with quantum numbers $0(0^+)$ exhibits a repulsive interaction. In contrast, both the $0(0^{-+})$ and $0(1^{--})$ channels of the $\Lambda_c\bar{\Lambda}_c$ system are found to be attractive, indicating the possible existence of bound states, particularly in the $0(1^{--})$ channel. In addition, there is a distinct mass splitting between the two channels that is attributed to the spin-spin terms in the TPE. Furthermore, we consider the SU(3) effects induced by the TKE contribution and refit the $C_a$ accordingly. Although both channels are still attractive in the SU(3) case, only the $0(1^{--})$ bound state remains. 

This work provides a consistent and comparative description of the $\Lambda_c\Lambda_c$ and $\Lambda_c\bar{\Lambda}_c$ interactions within a unified framework constrained by LQCD data. The predicted attractive potentials in the $\Lambda_c\bar{\Lambda}_c$ system, along with the possibility of bound states, as well as the predicted splitting brought by the TPE, motivate future experimental searches and more refined LQCD simulations aimed at exploring these exotic states.

\section*{ACKNOWLEDGMENTS}

Z. L. thanks Jun-Zhang Wang, Ri-Qing Qian, and Tian-Cai Peng for their valuable discussions. This work is supported by the Natural Science Foundation of Gansu Province (No. 26RCKA012, No. 25JRRA799, No. 22JR5RA389, and No. 22JR5RA171), the National Natural Science Foundation of China under Grants No. 12335001, No. 12247101, No. 12465016, and No. 12005168, the ``111 Center" under Grant No. B20063, the Fundamental Research Funds for the Central Universities (Grant No. lzujbky-2023-stlt01), and Lanzhou City High-Level Talent Funding.

\section*{DATA AVAILABILITY}

The data that support the findings of this article are
openly available \cite{Xing:2025uai, liu_2026_22110764}.


\bibliography{ref}

\end{document}